\documentclass[a4paper,12pt]{article}
\usepackage[utf8]{inputenc}
\usepackage[margin=1in]{geometry}
\usepackage{mathtools}
\usepackage{amssymb}
\usepackage{amsfonts}
\usepackage[footnotesize]{caption}
\usepackage[font=scriptsize]{subcaption}
\usepackage{color}
\usepackage{braket}
\usepackage{multirow}
\usepackage{relsize}
\usepackage{booktabs}
\usepackage{dcolumn}

\newcommand{\overbar}[1]{\mkern 1.5mu\overline{\mkern-1.5mu#1\mkern-1.5mu}\mkern 1.5mu}
\newcommand{\pmatr}[1]{\begin{pmatrix} #1 \end{pmatrix}}

\begin{document}

\begin{titlepage}
\begin{center}
{\bf\Large 
Accidental Peccei-Quinn Symmetry from Discrete Flavour Symmetry and Pati-Salam
} \\[12mm]
Fredrik~Bj\"{o}rkeroth$^{\star}$%
\footnote{E-mail: {\tt fredrik.bjorkeroth@lnf.infn.it}},
Eung~Jin~Chun$^{\dagger}$%
\footnote{E-mail: \texttt{ejchun@kias.re.kr}},
Stephen~F.~King$^{\ddagger}$%
\footnote{E-mail: \texttt{king@soton.ac.uk}},
\\[3ex]

\vspace*{0.50cm}
\centerline{$^{\star}$ \it INFN Laboratori Nazionali di Frascati, Via E. Fermi 40, 00044 Frascati, Italy}
\vspace*{0.2cm}
\centerline{$^{\dagger}$ \it 
Korea Institute for Advanced Study, Seoul 02455, Korea}
\vspace*{0.2cm}
\centerline{$^{\ddagger}$ \it
School of Physics and Astronomy, University of Southampton,}
\centerline{\it SO17 1BJ Southampton, United Kingdom }
\vspace*{1.20cm}
\end{center}

\begin{abstract}
{\noindent
We show how an accidental $U(1)$ Peccei-Quinn (PQ) symmetry can arise from a discrete $A_4$ family symmetry combined with a discrete flavour symmetry $ \mathbb{Z}_3 \times \mathbb{Z}_5^2 $, in a realistic Pati-Salam unified theory of flavour.
Imposing only these discrete flavour symmetries, the axion solution to the strong $ CP $ problem is protected from PQ-breaking operators to the required degree.
A QCD axion arises from a linear combination of $ A_4 $ triplet flavons, which are also responsible for fermion flavour structures due to their vacuum alignments. 
We find that the requirement of an accidental PQ symmetry arising from a discrete flavour symmetry constrains the form of the Yukawa matrices, providing a link between flavour and the strong $ CP $ problem. 
Our model predicts specific flavour-violating couplings of the flavourful axion and thus puts a strong limit on the axion scale from kaon decays.
}
\end{abstract}
\end{titlepage}

\section{Introduction}

Perhaps the best explanation for why $CP$ violation does not appear in strong interactions is to postulate a Peccei-Quinn (PQ) symmetry: a QCD-anomalous global $U(1)$ symmetry which is broken spontaneously, leading to a pseudo-Goldstone boson called the QCD axion \cite{pqww}.  
Typically, the PQ symmetry is realized by introducing heavy vector-like quarks (the KSVZ model) \cite{ksvz} or by extending the Higgs sector (the DFSZ model) \cite{dfsz}.
The QCD axion is also a good candidate for dark matter \cite{axionDM} within the allowed window of the axion (or PQ symmetry-breaking) scale $f_a=10^{9-12}$ GeV \cite{axion-review}.

It has also been realised that the PQ axion need not emerge from an exact global $U(1)$ symmetry, but could result from some discrete symmetry or continuous gauge symmetry leading to an accidental global $U(1)$ symmetry.
Considering the observed accuracy of strong-$CP$ invariance, it is enough to protect the PQ symmetry up to some higher-dimensional operators \cite{holman92}. 
In this regard, it is appealing to consider an approximate PQ symmetry gauranteed by discrete (gauge) symmetries \cite{chun92}.
Alternatively, an attempt to link PQ symmetry protected by continuous gauge symmetries to the flavour problem was made in \cite{cheung10}.%
\footnote{
	An approach to protecting the PQ symmetry to arbitrary order, based on gauged $ SU(N) \times SU(N) $, was recently proposed in \cite{DiLuzio:2017tjx}.  	
}

Despite being the leading candidate for a resolution to the strong $CP$ problem, the origin of the PQ symmetry and its possible connection with other aspects of physics remains unclear.
It is possible that PQ symmetry is related to flavour symmetries, which are a compelling proposal for understanding the origin of the fermion masses and mixing. 
Indeed it has already been proposed that the PQ symmetry arises from flavour symmetries  \cite{wilczek82}, linking the axion scale to the flavour symmetry-breaking scale, and various attempts have been made to incorporate such a flavourful PQ symmetry as a part of such \emph{continuous} flavour symmetries \cite{pq-flavour,ema16}. 
The resultant axion is sometimes dubbed a ``flaxion'' or ``axiflavon''.
We shall refer to it as a ``flavourful axion''.

In recent years there has been considerable work on \emph{discrete flavour symmetries} applied to understanding lepton -- especially neutrino -- masses and mixing parameters \cite{flavourmodelreviews}.
Motivated by this, we wish to put forward a new idea, namely that the PQ symmetry could arise accidentally from such 
\emph{discrete flavour symmetries} \cite{flavourmodelreviews}.%
\footnote{
	This idea should not be confused with alternatives to PQ symmetry, such as Nelson-Barr type resolutions to the strong $ CP $ problem \cite{nelsonbarroriginal}, or GUT models where specific Yukawa structures have been proposed \cite{nelsonbarrmodels}.
}
In order to include the quarks, as is necessary to resolve the strong $ CP $ problem, we shall combine discrete flavour symmetries with unified gauge theories where both quarks and leptons appear on an equal footing \cite{flavourmodelreviews}.
However for a discrete flavoured grand unified theory (GUT) based on $SO(10)$ \cite{Bjorkeroth:2015uou}, the flavour symmetry is often partly broken at the GUT scale, and it is hard to accommodate the traditional axion window below $10^{12}$ GeV.

In this work, we explore the possibilities of realizing an approximate or accidental PQ symmetry starting from a discrete flavour symmetry which controls both quarks and leptons via a Pati-Salam unification,
which allows a lower flavour breaking scale \cite{king14}.
The idea is that the discrete flavour symmetry and the resulting accidental PQ symmetry are both spontaneously broken by flavons at around $10^{11}$ GeV, leading to the observed flavour structure as well as the (approximate) QCD axion at the same time.
We demonstrate that a QCD axion at the correct scale can be achieved in a variant of the flavoured Pati-Salam (PS) model \cite{king14} employing an $ A_4 \times \mathbb{Z}_5 $ flavour symmetry, referred to as the ``A to Z'' model. 
We also show such a model is compatible with current quark and lepton mass and mixing data.

The layout of the remainder of the paper is as follows.
In Section~\ref{sec:model} we present the modified A to Z model: its field content, symmetries, and the superpotential responsible for flavour structures.
In Section~\ref{sec:strongcp} we show how it solves the strong $ CP $ problem, taking into account also higher-order corrections.
Section~\ref{sec:numerics} details the fermion mass and Yukawa matrices, and we perform a fit of the model to experimental data. 
Flavour constraints on the PQ-breaking scale are also considered in Section~\ref{sec:axionflavour}.
Section~\ref{sec:conclusion} concludes.

\section{The A to Z model}
\label{sec:model}

In this section we first give an overview of the original model, then propose a modification of it which is suitable for 
solving the strong $CP$ problem via an accidental global $U(1)_{PQ}$ symmetry emerging from an extended discrete flavour symmetry.

\subsection{Overview of the original model}

We here present the main features of the original A to Z model first introduced in \cite{king14}, before going on to introduce the modifications necessary to fully realise the automatic $ U(1)_{PQ} $ symmetry to the required accuracy. 
The model is based on the PS gauge group,
\begin{equation}
	G_{PS}=SU(4)_C \times SU(2)_L \times SU(2)_R, 
\end{equation}
which unifies left-handed (L) quarks and leptons, $F_i(4,2,1)$ and charge-conjugated right-handed (R) quarks and leptons $F^c_i(\bar{4},1,2)$, interpreting lepton number as a fourth colour, and where $i=1,2,3$ is a family index. 
In order to unify the left-handed families, it postulates an $ A_4 $ non-Abelian discrete flavour symmetry.
All left-handed SM chiral fields are united in an $ A_4 $ triplet $ F(3,4,2,1) $, under $A_4\times G_{PS}$, while right-handed fields are contained in three $ F^c_i (1,\bar{4},1,2)$, one for each generation.
$ F $ couples to $ A_4 $ triplet flavons $ \phi^u_{1,2} (3,1,1,1)$, $ \phi^d_{1,2} (3,1,1,1)$, under $A_4\times G_{PS}$, whose vacuum expectation values (VEVs) are aligned in particular directions according to constrained sequential dominance (CSD).
More precisely, the model realises the CSD(4) alignments along the $A_4$ triplet directions,
\begin{equation}
	\phi^u_1 \propto (0,1,1), \quad \phi^u_2 \propto (1,4,2), \quad
	\phi^d_1 \propto (1,0,0), \quad \phi^d_2 \propto (0,1,0),
\label{eq:csd4vevs}
\end{equation}
first explored in \cite{csd4}.
The fermion Yukawa matrices arise from non-renormalisable terms of the generic form $ (F \cdot \phi) h F^c $, where $ h $ denotes a Higgs superfield; these terms are realised by a renormalisable superpotential involving messengers, generally denoted $ X $, which are integrated out below the GUT scale.

As discussed in more detail below, CSD(4) leads to up-type quark and neutrino Yukawa matrices ($ Y^u $ and $ Y^\nu $, respectively) of the form (in LR convention)
\begin{equation}
	Y^u = Y^\nu \sim \pmatr{0 & b & 0 \\ a & 4b & 0 \\ a & 2b & c},
\end{equation}
while the down-type quark and charged lepton Yukawa matrices ($ Y^d $ and $ Y^e $) are approximately diagonal, owing to flavons $ \phi^d_{1,2} $ whose VEVs are aligned in the $ (1,0,0) $ and $ (0,1,0) $ directions.
The third family couplings arise from a renormalisable interaction $ F h_3 F^c_3 $ where $ h_3 $ is an $ A_4 $ triplet.

These mass matrix structures yield predictions for quark and lepton mixing, including a natural prediction for the Cabibbo angle $ \theta^q_{12} \sim 1/4 $ and a neutrino reactor angle $ \theta^\ell_{13} \sim 9.5^\circ $, subject to small corrections from the off-diagonal elements of $ Y^{d,e} $.
The original prediction for $ \theta^\ell_{13} $ agreed well with experiment at the time, while more recent global fits (e.g. \cite{nufit3}) prefer smaller $ \theta^\ell_{13} \approx 8.5^\circ $. 
In the modified theory presented in this paper, the precise structure of the Yukawa matrices differ slightly; we show that it can accommodate current experimental data. 
The Higgs sector is discussed in some detail in \cite{king14}, with an explicit mechanism given for spontaneous breaking of PS $ \to $ SM.
As all Higgs fields are understood to be neutral under the accidental PQ symmetry, these results remain intact, and will not be discussed further here; we refer interested readers to the original paper.

The original A to Z model also involves a discrete $\mathbb{Z}_5 $ flavour symmetry.
We will show below that the model very nearly possesses an accidental global $ U(1) $ symmetry, suggesting that a PQ solution to the strong $ CP $ problem may be realised in this model with only minor modifications.
The most significant addition is a discrete $ \mathbb{Z}_3 $ symmetry under which matter, flavons, and messengers are charged.
This ensures certain (previously allowed) operators in the renormalisable superpotential, which explicitly break the accidental $ U(1)_{PQ} $, are forbidden.
An additional $ \mathbb{Z}_5^\prime $ is necessary to forbid higher-order terms up to a required order. 
Finally, the mechanism originally proposed to drive the flavon VEVs to a particular scale is not suitable since it does not respect the accidental $ U(1)_{PQ} $. 
Instead we shall discuss an alternative mechanism which preserves $ U(1)_{PQ} $, and discuss its implications for flavour.

\subsection{The modified model}

\begin{table}[htb]
\centering
\begin{tabular}{|c|cccccc||c|}
\hline
\rule{0pt}{3ex}%
Field & $G_{PS}$ & $A_4$ & $\mathbb{Z}_5 $ & $ \mathbb{Z}_3 $ & $\mathbb{Z}'_5 $ & $R$ & $ U(1)_{PQ} $ \\[0.3ex]
\hline\hline
\rule{0pt}{2.5ex}%
$ F $ & $ (4,2,1) $ & 3 & 1 & 1 & 1 & 1 & 0 \\
$ F^c_{1,2,3} $ & $(\overbar{4},1,2)$ & 1 & $\alpha, \alpha^3, 1$ & $ \beta, \beta^2, 1 $ & $ \gamma^3, \gamma^4, 1 $ & 1 & $ -2, -1, 0 $ \\[0.3ex]
\hline
\rule{0pt}{2.5ex}%
$\overbar{H^c}$ & $(4,1,2)$ & 1 & 1 & 1 & 1 & 0 & 0  \\
$H^c$  & $(\overbar{4},1,2)$ & 1 & 1 & 1 & 1 & 0 & 0 \\[0.3ex]
\hline
\rule{0pt}{2.5ex}%
${\phi}^{u}_{1,2}$  & $(1,1,1)$ & 3 & $\alpha^4,\alpha^2$  & $ \beta^2, \beta$ & $ \gamma^2, \gamma$ & 0 & $ 2, 1 $\\
${\phi}^{d}_{1,2}$  & $(1,1,1)$ & 3 & $\alpha^3,\alpha$ & $ \beta^2, \beta$ & $ \gamma^2, \gamma$ & 0 & $ 2, 1 $ \\[0.3ex]
\hline
\rule{0pt}{2.5ex}%
$h_3$  & $(1,2,2)$  & 3 & 1 & 1 & 1 & 0 & 0  \\
$h_u$  & $(1,2,2)$  &$1''$ & $\alpha$ & 1 & 1 & 0 & 0 \\
$h^u_{15}$  & $(15,2,2)$  &1 & $\alpha$ & 1 & 1 & 0 & 0  \\
$h_d$  & $(1,2,2)$  & $1'$ & $\alpha^3$ & 1 & 1 & 0 & 0\\
$h^d_{15}$  & $(15,2,2)$  &$1'$ & $\alpha^4$ & 1 & 1 & 0 & 0 \\[0.3ex]
\hline
\rule{0pt}{2.5ex}%
$\Sigma_u$  & $(1,1,1)$  &$ 1'' $ & $\alpha$ & 1 & 1 & 0 & 0  \\
$\Sigma_d$  & $(1,1,1)$  &$ 1' $ & $\alpha^3$  & 1 & 1 & 0 & 0 \\
$\Sigma^d_{15}$  & $(15,1,1)$  &$ 1' $ & $\alpha^2$  & 1 & 1 & 0 & 0  \\[0.3ex]
\hline
\rule{0pt}{2.5ex}%
$\xi$ & $(1,1,1)$  &$ 1$ &  $\alpha^4$ & $\beta^2$ & $ \gamma^2 $ & 0 & 2  \\[0.3ex]
\hline
\rule{0pt}{2.5ex}%
$X_{F''_{1,3}}$ & $(4,2,1)$  &$ 1''$&  $\alpha$,$\alpha^3$ & $ \beta^2,\beta $ & $ \gamma^2, \gamma $ & 1 & $ 2, 1 $ \\
$X_{F'_{1,3}}$ & $(4,2,1)$  &$1'$&  $\alpha$,$\alpha^3$  & $ \beta,\beta^2 $ & $ \gamma, \gamma^2 $ & 1 & $ 1, 2 $\\
$X_{\overbar{F_i}}$ & $(\overbar{4},2,1)$  & 1 &  $\alpha^i$ & $\beta,\beta,\beta^2,\beta^2$ & $ \gamma^3, \gamma^3, \gamma^4, \gamma^4 $ & 1 & $ -2, -2, -1, -1 $  \\
$X_{\xi_{i}}$ & $(1,1,1)$  &$ 1$ &  $\alpha^i$  & $ \beta,\beta,\beta^2,\beta^2,1 $ & $ \gamma^3,\gamma,\gamma^4,\gamma^2,1 $ & 1 & $  -2, 1, -1, 2, 0 $ \\[0.3ex]
\hline\hline
\rule{0pt}{2.5ex}%
$\overbar{\phi}_{1,2}^u$ & $(1,1,1)$ & 3 & $ \alpha, \alpha^3 $ & $ \beta, \beta^2 $ & $ \gamma^3, \gamma^4 $ & 0 & $ -2, -1 $ \\
$\overbar{\phi}_{1,2}^d$ & $(1,1,1)$ & 3 & $ \alpha^2, \alpha^4 $ & $ \beta, \beta^2 $ & $ \gamma^3, \gamma^4 $ & 0 & $ -2, -1 $ \\
$ \bar{\xi} $ & $ (1,1,1) $ & 1 & $ \alpha $ & $ \beta $ & $ \gamma^3 $ & 0 & $-2$ \\[0.3ex]
\hline
\end{tabular}
\caption{
	The basic Higgs, matter, flavon and messenger content of the model.
	$\alpha = \gamma = e^{2\pi i /5}$, and $\beta=e^{2\pi i/3}$.
	$R$ is a supersymmetric $R$-symmetry.
	We emphasise that $U(1)_{PQ}$ is a resulting approximate PQ symmetry which is not imposed 
	directly, but emerges as an accidental result of the discrete flavour symmetry.
}
\label{tab:fields}
\end{table}

The field content of the modified A to Z model are given in Table~\ref{tab:fields}, showing their charges under gauge and discrete symmetries, as well as the inferred charges under the accidental $ U(1)_{PQ} $, although we emphasise that this symmetry is not enforced but arises as a consequence of the discrete flavour symmetry.
We split the renormalisable superpotential into several parts,
\begin{equation}
	W = W_F + W_\mathrm{Maj} + W_\mathrm{driving} + W_H .
\end{equation}
The Higgs part $ W_H $ plays no role in the solution to the strong $ CP $ problem since  the Higgs fields $ h_i $, $ H^c $ are invariant under $ U(1)_{PQ} $ and have zero inferred PQ charges.
In fact the only fields whose scalar components get VEVs and which carry PQ charge are the flavons, which are therefore solely responsible for PQ-symmetry breaking.
The driving superpotential $ W_\mathrm{driving} $, which sets the scale of flavon VEVs, will be discussed shortly.
The ``fermion'' part $ W_F $, containing the couplings of $ F $ and $ F^c_i $ to flavons and $ X $ messengers as well as messenger couplings to $ \Sigma $ fields, is given by
\begin{equation}
\begin{split}
	W_F &=
	F h_3 F^c_3 
	\\ & \qquad
	+ X_{\overbar F_1} \phi^u_1 F 
	+ X_{\overbar F_2} \phi^d_1 F
	+ X_{\overbar F_3} \phi^u_2 F
	+ X_{\overbar F_4} \phi^d_2 F 
	\\ & \qquad
	+ X_{F'_1} h_u F^c_2
	+ X_{F'_3} h_u F^c_1
	+ X_{F''_1} h_d F^c_1
	% + X_{F''_1} h^d_{15} F^c_3 % Forbidden by Z3!
	+ X_{F''_3} h^d_{15} F^c_2 
	\\ & \qquad
	+ X_{F'_1}\Sigma_u X_{\overbar F_3}
	+ X_{F'_3}\Sigma_u X_{\overbar F_1}
	+ X_{F''_1}(
	\Sigma_d X_{\overbar F_1} + \Sigma^d_{15}  X_{\overbar F_2}
	)
	+ X_{F''_3}\Sigma_d X_{\overbar F_4} ,
\end{split}
\end{equation}
The ``Majorana'' part $ W_\mathrm{Maj} $ gives the right-handed neutrino mass matrix, and is given by
\begin{equation}
\begin{split}
	W_\mathrm{Maj} &=
	X_{\xi_4} \overbar{H^c} F^c_1 
	+ X_{\xi_2} \overbar{H^c} F^c_2 
	+ X_{\xi_5} \overbar{H^c} F^c_3 
	\\ & \qquad
	+ \Lambda X_{\xi_1} X_{\xi_4} 
	+ \xi X_{\xi_1} X_{\xi_5} 
	+ \Lambda X_{\xi_2} X_{\xi_3} 
	+ \xi X_{\xi_3} X_{\xi_3}
	+ \Lambda X_{\xi_5} X_{\xi_5} .
\end{split}
\end{equation}
\begin{figure}[htbp]
\centering
	\includegraphics[scale=0.8]{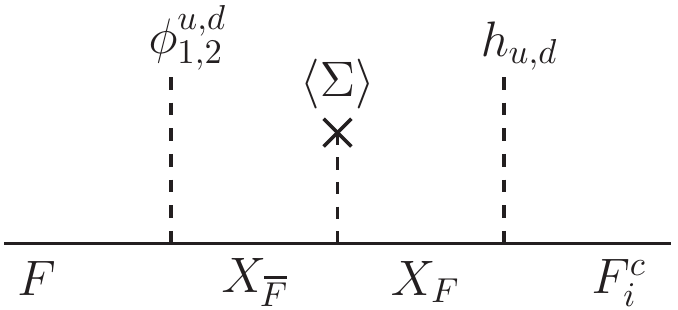}
\caption{Generic diagram representing all the effective fermion Yukawa terms $ W_F^\mathrm{eff} $.}
\label{fig:ffc}
\end{figure}
\begin{figure}
\centering
	\includegraphics[scale=0.8]{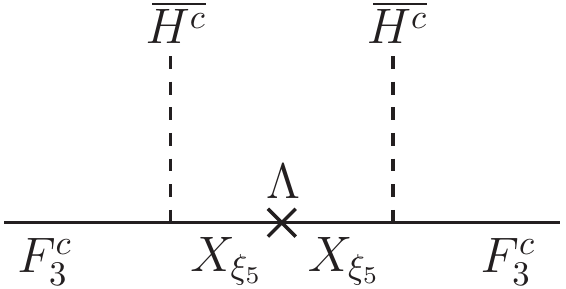}\qquad
	\includegraphics[scale=0.8]{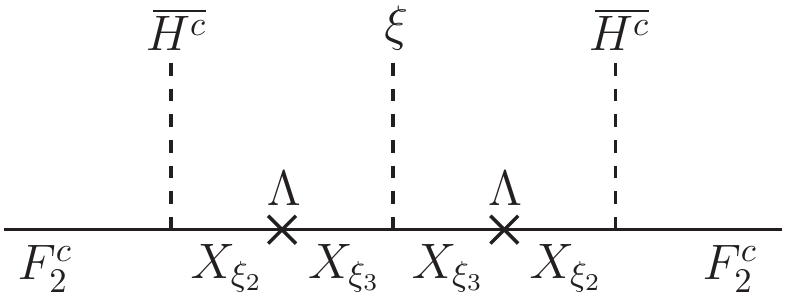}
\caption{Diagrams giving effective Majorana terms in $ W^\mathrm{eff}_\mathrm{Maj} $.
The Majorana terms proportional to $ F^c_1 F^c_1 $ and $ F^c_1 F^c_3 $, not shown, are constructed in a similar way.
}
\label{fig:fcfc}
\end{figure}
After integrating out $ X $ messengers, we obtain the effective superpotentials, which also preserve $ U(1)_{PQ} $,
\begin{equation}
\begin{split}
	W_F^\mathrm{eff} &= 
	(F \cdot h_3) F^c_3
	+ \frac{(F \cdot \phi_1^u) h_u F^c_1}{\braket{\Sigma_u}}
	+ \frac{(F \cdot \phi_2^u) h_u F^c_2}{\braket{\Sigma_u}}
	\\ & \qquad
	+ \frac{(F \cdot \phi_1^d) h_d F^c_1}{\braket{\Sigma^d_{15}}}
	+ \frac{(F \cdot \phi_2^d) h^d_{15} F^c_2}{\braket{\Sigma_d}}
	+ \frac{(F \cdot \phi_1^u) h_d F^c_1}{\braket{\Sigma_d}} ,
	\\
	W_\mathrm{Maj}^\mathrm{eff} &=
	\frac{\overbar{H^c}\overbar{H^c}}{\Lambda} \left(
	\frac{\xi^2}{\Lambda^2}  F^c_1 F^c_1
	+ \frac{\xi}{\Lambda} F^c_2 F^c_2
	+ F^c_3 F^c_3 
	+ \frac{\xi}{\Lambda} F^c_1 F^c_3
	\right) .
\end{split}
\label{eq:Wnonrenorm}
\end{equation}
Figures~\ref{fig:ffc} and \ref{fig:fcfc} show how non-renormalisable terms arise in $ W_F^\mathrm{eff} $ and $ W_\mathrm{Maj} ^\mathrm{eff}$, respectively, from diagrams involving $ X $ messengers which receive large masses either from $\Sigma$ 
fields which get VEVs, or have direct heavy masses $\Lambda$. 

It is worth recalling that, 
in the original model, the superpotential allowed a term $ X_{F_1^{\prime\prime}} h_{15}^d F^c_3 $, which gave rise to two effective terms
\begin{equation}
	W \supset 
	\frac{(F \cdot \phi_1^d) h_{15}^d F^c_3}{\braket{\Sigma_{15}^d}} 
	+ \frac{(F \cdot \phi_1^u) h_{15}^d F^c_3}{\braket{\Sigma_d}},
\label{eq:forbiddenterms}
\end{equation}
which populated the (1,3) and (2,3) elements of $ Y^d $ and $ Y^e $.
These terms, which would violate the accidental PQ symmetry,
are now both forbidden by the extended flavour symmetry.

\subsection{Driving sector}

Yukawa textures are controlled by vacuum alignments of triplet flavons. 
In addition, we wish to drive the flavon VEVs to a particular scale.
One possible mechanism which achieves this while preserving the strong $ CP $ solution is to introduce ``conjugate'' flavons, labelled $ \overbar{\phi}_{1,2}^{u,d} $, $ \bar{\xi} $, which have charges opposite to $ \phi_{1,2}^{u,d} $ and $ \xi $, respectively, including under $ U(1)_{PQ} $.
We are able to drive flavon VEVs to a scale $ M $ via a superpotential
\begin{equation}
	W_\mathrm{driving} = 
	P_{1,2}^{u,d} \left( \overbar{\phi}_{1,2}^{u,d} \phi_{1,2}^{u,d} - M^2 \right) 
	+ P_\xi \left( \bar{\xi} \xi - M^2 \right) ,
\label{eq:Wdriving}
\end{equation}
with each of the five $ \overbar{\phi}\phi $ or $ \bar{\xi}\xi $ pairs driven by the $ F $-term equation of the corresponding driving field $ P $, which has $ R = 2 $.

It is worth recalling that 
the original A to Z model permitted bilinear flavon terms $ \phi_1^u \phi_2^d $ and $ \phi_2^u \phi_1^d $, invariant under all discrete symmetries.
These coupled to driving fields $ P_{ij} $, giving rise to driving terms $ W \supset P_{12}(\phi_1^u \phi_2^d - M_{12}^2) + P_{21}(\phi_2^u \phi_1^d - M_{21}^2) $.
However these bilinears have total charge 3 under $ U(1)_{PQ} $, breaking it explicitly.
As such, the original mechanism is not compatible with an accidental PQ symmetry.
This could have been remedied by removing the driving fields $ P_{ij} $ from the theory, but this would have left the question of how flavons acquire non-zero VEVs.%
\footnote{
	It is possible that the flavon VEVs are driven by another mechanism entirely, without introducing new conjugate flavons and driving fields.
	For instance, in a theory of radiative SUSY breaking, loop corrections may drive the mass parameter to a negative value, leading to non-zero values at the potential minimum \cite{Howl:2009ds}.
}
In the previous discussion we have doubled the flavon sector by introducing $\overbar{\phi}_{1,2}^{u,d}$ with opposite effective PQ charges
as shown in Table~\ref{tab:fields} and Eq.~\ref{eq:Wdriving}.

\section{Solution to the strong \boldmath{$CP$} problem}
\label{sec:strongcp}

\subsection{Accidental QCD axion from flavon fields}

The modified model described in the previous section 
has the necessary ingredients to realise a PQ symmetry (leading to a QCD axion): a global chiral $ U(1) $ symmetry, which  is  spontaneously broken when $ A_4 $-triplet flavon fields acquire non-zero VEVs, and a colour anomaly, which is ensured by the (standard) left- and right-handed fermions (contained in $ F $ and $ F^c_i $, respectively) transforming differently under the $ U(1)_{PQ} $.
 
We now show in more detail how the A to Z model solves the strong $ CP $ problem. 
The QCD axion $a$ arises as a combination of the phases of the flavons $ \phi^{u,d}_{1,2} $ and $\xi$,
\begin{equation} \label{a-aphi}
	a= \frac{1}{v_{PQ}} \sum_\varphi x_\varphi v_\varphi a_\varphi %/ v_{PQ}
\end{equation}
where $x_\varphi$ denotes the PQ charge of a flavon $\varphi$ (that is, $x_{\phi^{u,d}_1}=2$, $x_{\phi^{u,d}_2}=1$ and $x_\xi=2$), $v_\varphi$ is the VEV of $\varphi$ (i.e. $\braket{\varphi^\dagger \varphi} = v^2_\varphi/2$),%
\footnote{
	More precisely for $ A_4 $ triplets, $ v_\varphi/\sqrt{2} $ is the constant of proportionality of the alignments described in Eq.~\ref{eq:csd4vevs}.
}
$a_\varphi$ is the phase field of $\varphi$, and $v_{PQ}=\sqrt{ \sum_\varphi x^2_\varphi v^2_\varphi}$.
In order to get the correct scale of the Yukawa couplings for the first and second families of fermions, we require these to acquire VEVs of $ \mathcal{O}(10^{11}) $ GeV, which is the desired scale of PQ breaking.  

The QCD anomaly number $N_a$ of the PQ symmetry is determined by the sum of the PQ charges of the fermion fields $F, F^c_i$ and messengers $X$. 
As the $ X $ are vector-like, they do not contribute to $N_a$ and thus we have $N_a \equiv |6 x_F + 2\sum_i x_{F^c_i}|=6$. 

The axion-gluon-gluon coupling is:
\begin{equation}
\mathcal{L}_{agg} = \frac{\alpha_s}{8\pi} \frac{a}{f_a} G^a_{\mu\nu} \tilde G_a^{\mu\nu}
\end{equation}
where  $f_a=v_{PQ}/N_a$, which leads to the axion mass $m_a \approx m_\pi f_\pi/f_a$.
Although our model differs from the DFSZ model in that the standard Higgs doublets are not charged under the PQ symmetry, the QCD anomaly is the same and the axion phenomenology is very similar. 
A crucial difference comes from flavour-violating axion couplings, as will be discussed in a later section. 
Since our model has a number of discrete symmetries which are supposed to be broken at around $10^{11}$ GeV, it is vulnerable to a dangerous domain wall problem. 
This can in principle be evaded if symmetry breaking occurs during inflation. 
However we shall not discuss inflation here.

\subsection{Corrections from higher-order operators}

Flavour models based on non-Abelian discrete symmetries typically require the inclusion of one or more Abelian ``shaping symmetry'' groups which ensure the superpotential includes only desirable terms at low order, leading to predictive mass structures \cite{flavourmodelreviews}. 
The $ \mathbb{Z}_3 $ symmetry ensures a $ U(1)_{PQ} $ in the renormalisable theory involving the flavons $ \phi_{1,2}^{u,d} $.
However, high-dimensional operators  involving powers of the PQ scalar field(s) (in this model, flavons) explicitly violating $ U(1)_{PQ} $ may be present to shift the axion potential away from its $ CP $-conserving minimum \cite{holman92}.

We first consider higher-dimensional terms of the form 
\begin{equation}
	\frac{\{\phi\}^n}{M_P^n} W,
\label{eq:effectivephiterms}
\end{equation}
where $ \{ \phi \}^n $ denotes any combination of flavons $ \phi_{1,2}^{u,d} $ (or their $ \overbar{\phi} $ counterparts) which
are allowed by the discrete flavour symmetry and $R$-symmetry under consideration, but which do not respect the accidental PQ symmetry.

In the context of supergravity, supersymmetry breaking generically leads to the VEV $\braket{W} \sim m_{3/2} M_P^2$ where $ m_{3/2} $ is the gravitino mass.
The operator in Eq.~\ref{eq:effectivephiterms} then generates a PQ-breaking axion mass contribution
\begin{equation}
 	m_\ast^2 \sim m^2_{3/2} \frac{v_{PQ}^{n-2}}{M_P^{n-2}} .
\end{equation} 
To preserve the axion solution to the strong $CP$ problem, we require $ m_\ast^2/m_a^2 < 10^{-10} $,
where the standard axion mass due to QCD instantons is given by $ m_a^2 \approx m_\pi^2 f_\pi^2/f_a^2 $ \cite{holman92}.
Taking $ v_{PQ} \sim 10^{11} $ GeV and  $ M_P = 2\times 10^{18} $ GeV, the above condition is satisfied with $n>7$.

We must therefore forbid all flavon combinations up to $ n = 7 $, or equivalently superpotential terms with $ D = 10 $ (since $ \mathrm{dim}(W) = 3 $), to ensure the solution to the strong $ CP $ problem  protected to sufficient order.
This cannot be achieved by $ \mathbb{Z}_5 \times \mathbb{Z}_3 $ alone.
This requires the additional $ \mathbb{Z}_5^\prime $ symmetry, which protects the PQ solution, but largely does not alter the renormalisable superpotential or flavour predictions.
It is worth noting, however, that if we assume flavon VEVs are driven by the superpotential proposed in Eq.~\ref{eq:Wdriving}, the gauge and $ A_4 \times \mathbb{Z}_5 \times \mathbb{Z}_3 $ symmetries would also permit renormalisable interactions involving the conjugate flavons $ \overbar{\phi}_{1,2}^{u,d} $ with matter of the form $ X_{\overbar{F}} \overbar{\phi} F $, which violate $ U(1)_{PQ} $. However the 
$ \mathbb{Z}_5^\prime $ ensures these terms are forbidden.

\section{Mass structures and numerical fit}
\label{sec:numerics}

In this section we show that the modified model provides a realistic description of all quark and lepton
masses and mixing parameters.
From the terms in Eq.~\ref{eq:Wnonrenorm}, when the flavons acquire CSD(4) VEVs, we arrive at the Yukawa matrices
\begin{equation} \label{Yuk}
\begin{split}
	Y^u = Y^\nu &= \pmatr{
		0 & b & \epsilon_{13} c \\
		a  & 4 b & \epsilon_{23} c \\
		a  & 2 b & c \\
	}, 
	\quad
	Y^d = \pmatr{
		y_d^0  & 0 & 0 \\
		B y_d^0  & y_s^0  & 0 \\
		B y_d^0  & 0 & y_b^0
	}, 
	\quad
	Y^e = \pmatr{
		-(y_d^0/3)  & 0 & 0 \\
		B y_d^0  & x y_s^0  & 0 \\
		B y_d^0  & 0 & y_b^0
	} ,
\end{split}
\end{equation}
where all parameters are in general complex.
The parameters $ a $, $ b $, $ y_d^0 $, and $ y_s^0 $ are proportional to the VEVs of flavons $ \phi_1^u $, $ \phi_2^u $, $ \phi_1^d $, and $ \phi_2^d $, respectively, while $ c $, $ y_b^0 $ derive from the renormalisable coupling $ F h_3 F^c_3 $. 
Elements proportional to $ B $ arise from the term $ F \phi_1^u h_d F^c_1 $; $ B $ is expected to be an $ \mathcal{O}(1) $ number.
$ \epsilon_{13,23} $ parametrise a small admixture of $ h_3 $ into $ h_u $, which is not specified by the model, while
$ x $ is related to the mixing between doublets within $ h_d $ and $ h_{15}^d $, and is taken to be real.
One overall phase in each Yukawa matrix is unphysical; we may choose $ c $ and $ y_b^0 $ to be real without loss of generality.
The right-handed neutrino mass matrix $ M_R $ is given by
\begin{equation}
	M_R = \pmatr{
		M_{1} & 0 & M_{13} \\
		0 & M_{2} & 0 \\
		M_{13} & 0 & M_{3}
	} .
\end{equation}
The neutrino matrix after seesaw is
\begin{equation}
	m^\nu = 
	m_a \pmatr{0&0&0\\0&1&1\\0&1&1}
	+ m_b e^{i \eta} \pmatr{1&4&2\\4&16&8\\2&8&4}
	+ m_c e^{i \xi} \pmatr{0&0&0\\0&0&0\\0&0&1},
\end{equation}
where we have neglected the contributions from $ \epsilon_{13,23} $, as they are small and appear only at $ \mathcal{O}(\epsilon^2) $.
We also neglect contributions from the off-diagonal elements $ M_{13} $ in $ M_R $.
The Yukawa matrices differ from those in the original model in two ways. 
Off-diagonal (1,3) and (2,3) elements of $ Y^d $ and $ Y^e $ are now zero, as the effective terms that produced these contributions (Eq.~\ref{eq:forbiddenterms}) are forbidden by the $ \mathbb{Z}_3 $ symmetry.
Moreover, all phases are now free, unlike in the original model where they were fixed to discrete multiples of $ 2\pi/5 $.
This phase fixing derived from the driving superpotential which, as noted above, is incompatible with a PQ solution.
In summary, the additional texture zeroes in $ Y^{d,e} $ increase overall predictivity, while the additional phase freedom decreases it.

\begin{table}[htb]
\centering
\begin{tabular}{ l c D{@}{\ \to \ }{6,6} c@{\hskip 3pt} c D{@}{\ \to \ }{6,6} }
\toprule
	\multirow{2}{*}{Observable}& \multicolumn{2}{c}{Data} && \multicolumn{2}{c}{Model} \\
\cmidrule{2-3} \cmidrule{5-6}
	& Central value & \multicolumn{1}{c}{1$\sigma$ range}  && Best fit & \multicolumn{1}{c}{Interval} \\
\midrule
	$\theta_{12}^\ell \, /^\circ$ 	& 33.57 & 32.81 @ 34.32 && 32.88 & 32.72 @ 34.23 \\ 
	$\theta_{13}^\ell \, /^\circ$ 	& 8.460 & 8.310 @ 8.610 && 8.611 & 8.326 @ 8.882 \\  
	$\theta_{23}^\ell \, /^\circ$ 	& 41.75 & 40.40 @ 43.10 && 39.27 & 37.35 @ 40.11 \\ 
	$\delta^\ell \, /^\circ$ 		& 261.0 & 202.0 @ 312.0 && 242.6 & 231.4 @ 249.9 \\
	$y_e$  $/ 10^{-5}$ 				& 1.004 & 0.998 @ 1.010 && 1.006 & 0.911 @ 1.015 \\ 
	$y_\mu$  $/ 10^{-3}$ 			& 2.119 & 2.106 @ 2.132 && 2.116 & 2.093 @ 2.144 \\ 
	$y_\tau$  $/ 10^{-2}$	 		& 3.606 & 3.588 @ 3.625 && 3.607 & 3.569 @ 3.643 \\ 
	$\Delta m_{21}^2 \, / 10^{-5} \, \mathrm{eV}^2 $ 
									& 7.510  & 7.330 @ 7.690 && 7.413 & 7.049 @ 7.762 \\
	$\Delta m_{31}^2 \, / 10^{-3} \, \mathrm{eV}^2 $ 
									& 2.524 & 2.484 @ 2.564 && 2.540 & 2.459 @ 2.616 \\
	$m_1$ /meV 						& & \multicolumn{1}{c}{} && 0.187 & 0.022 @ 0.234 \\ 
	$m_2$ /meV						& & \multicolumn{1}{c}{} && 8.612 & 8.400 @ 8.815 \\ 
	$m_3$ /meV 						& & \multicolumn{1}{c}{} && 50.40 & 49.59 @ 51.14 \\
	$\sum m_i$ /meV 				& & \multicolumn{1}{c}{$<$ 230} && 59.20 & 58.82 @ 60.19 \\
	$ \alpha_{21} $ 				& & \multicolumn{1}{c}{} && 10.4 & -38.0 @ 70.1 \\
	$ \alpha_{31} $ 				& & \multicolumn{1}{c}{} && 272.1 & 218.2 @ 334.0 \\
	$ m_{\beta\beta} $ /meV 		& & \multicolumn{1}{c}{} && 1.940 & 1.892 @ 1.998 \\
\bottomrule
\end{tabular}
\caption[]{
	Model predictions in the lepton sector, at the GUT scale.
	We set $\tan \beta = 5$, $ M_\mathrm{SUSY} = 1 $ TeV and $ \bar{\eta}_b = -0.24 $. 
	The model interval is a Bayesian 95\% credible interval. 
	The bound on $ \sum m_i $ is taken from \cite{planck}.
}
\label{tab:outputleptons}
\end{table}

\begin{table}[htb]
\centering
\begin{tabular}{ l c D{@}{\ \to \ }{6,6} c@{\hskip 3pt} c D{@}{\ \to \ }{6,6} }
\toprule
	\multirow{2}{*}{Observable} & \multicolumn{2}{c}{Data} && \multicolumn{2}{c}{Model} \\
\cmidrule{2-3} \cmidrule{5-6}
	& Central value & \multicolumn{1}{c}{$1\sigma$ range} && Best fit & \multicolumn{1}{c}{Interval} \\
\midrule
	$\theta_{12}^q \, /^\circ$ & 13.03 & 12.99 @ 13.07 && 13.04 & 12.94 @ 13.11 \\	
	$\theta_{13}^q \, /^\circ$ & 0.1471 & 0.1418 @ 0.1524 && 0.1463 & 0.1368 @ 0.1577 \\
	$\theta_{23}^q \, /^\circ$ & 1.700 & 1.673 @ 1.727 && 1.689 & 1.645 @ 1.753 \\	
	$\delta^q \, /^\circ$ & 69.22 & 66.12 @ 72.31 && 68.85 & 63.00 @ 75.24 \\
	$y_u \, / 10^{-6}$ & 2.982 & 2.057 @ 3.906 && 3.038 & 1.098 @ 4.957  \\	
	$y_c \, / 10^{-3}$ & 1.459 & 1.408 @ 1.510 && 1.432 & 1.354 @ 1.560 \\	
	$y_t$  			   & 0.544 & 0.537 @ 0.551 && 0.545 & 0.530 @ 0.558 \\
	$y_d \, / 10^{-5}$ & 2.453 & 2.183 @ 2.722 && 2.296 & 2.181 @ 2.966 \\	
	$y_s \, / 10^{-4}$ & 4.856 & 4.594 @ 5.118 && 4.733 & 4.273 @ 5.379 \\
	$y_b$   		   & 3.616 & 3.500 @ 3.731 && 3.607 & 3.569 @ 3.643 \\
\bottomrule
\end{tabular}
\caption[]{
	Model predictions in the quark sector at the GUT scale.
	We set $\tan\beta = 5$, $M_\mathrm{SUSY} = 1$ TeV and $\bar{\eta}_b = -0.24$. 
	The model interval is a Bayesian 95\% credible interval.
}
\label{tab:outputquarks}
\end{table}

\begin{table}[ht]
\centering
% \footnotesize
\renewcommand{\arraystretch}{1.1}
\begin{tabular}[t]{lr}
\toprule
	Parameter & Value \\ 
\midrule
	$a \, /10^{-5}$ & $1.246 \, e^{4.047i}$ \\
	$b \, /10^{-3}$ & $3.438 \, e^{2.080i}$ \\
	$c$ 			& $-0.545$ \\
	$y_d^0 \, /10^{-5}$ & $3.053 \, e^{4.816i}$ \\
	$y_s^0 \, /10^{-4}$ & $3.560 \, e^{2.097i}$ \\
	$y_b^0 \, /10^{-2}$ & $3.607$ \\
	$\epsilon_{13} \, /10^{-3}$ & $6.215 \, e^{2.434i}$ \\
	$\epsilon_{23} \, /10^{-2}$ & $2.888 \, e^{3.867i}$ \\
	$B$ & $10.20 \, e^{2.777i}$ \\ 
	$x$	& $5.880$ \\
\bottomrule
\end{tabular}
\hspace*{0.5cm}
\begin{tabular}[t]{lr}
\toprule
	Parameter & Value \\ 
\midrule
	$m_a$ /meV & $3.646$ \\
	$m_b$ /meV & $1.935$ \\
	$m_c$ /meV & $1.151$ \\
	$\eta$ & $2.592$ \\
	$\xi$ & $2.039$ \\
\bottomrule
\end{tabular}
\hspace*{0.5cm}
\caption{Best fit input parameter values. 
%% Taken from plain_1mil
} 
\label{tab:parameters}
\end{table}

The best fit of the model to quark \cite{antuschmaurer} and lepton \cite{nufit3} data is given in Tables~\ref{tab:outputleptons} and \ref{tab:outputquarks}. 
We have performed a Markov Chain Monte Carlo (MCMC) analysis to find the best fit to mass and mixing parameters at the GUT scale, and to estimate the predicted ranges of physical parameters.
The running of Yukawa couplings and quark mixing angles depends on various SUSY-breaking parameters, including $ \tan \beta $, the scale of SUSY breaking $ M_\mathrm{SUSY} $, and threshold corrections.
These have been studied in \cite{antuschmaurer}, which parametrise SUSY threshold effects in terms of several parameters $ \bar{\eta}_i $.
We assume a small correction corresponding to the choice $ \bar{\eta}_b = -0.24 $, which primarily affects the $ b $ quark Yukawa coupling, ensuring $ y_b \approx y_\tau $ at $ M_\mathrm{GUT} $ as predicted by the model.
All other threshold parameters are set to zero.
We assume no running in the PMNS parameters.
The intervals given correspond to so-called 95\% credible intervals, which correspond, in standard Bayesian formalism, to the regions in parameter space of highest posterior (probability) density (hpd).%
\footnote{
	This is analogous to, but should not be confused with, frequentist confidence intervals.
}

The fit gives minimum $ \chi^2 \approx 7.2 $.
Generically the model predicts a rather large Cabibbo angle $ \theta_{12}^q \sim 1/4 $, while $ \theta_{13,23}^q $ are expected to be much smaller.
There is sufficient parameter freedom to attain a fit to all quark mass and mixing parameters to within $ 1\sigma $ of their experimental best fits.
However in the lepton sector a perfect fit is never attained: the fit consistently predicts an atmospheric angle $ \theta_{23}^\ell \sim 38-39^\circ $, well below the current experimental value around $ 42^\circ $.
This is the dominant contribution to the $ \chi^2 $.
This agrees with earlier analytical results: as shown in \cite{csd4}, when the mass parameters $ m_{a,b} $ in $ m^\nu $ are fixed to give the correct neutrino mass-squared differences,%
\footnote{
	The third mass parameter, $ m_c $, has only a small effect in the strongly hierarchical regime preferred by the fit.
}
the mixing angles in CSD(4) are found to obey an approximate sum rule $ \theta^\ell_{23} \sim 45^\circ + \sqrt{2} \theta^\ell_{13} \cos \delta^\ell $, which implies small $ \theta^\ell_{23} $.
This prediction may be tested by increased precision in the measurement of the PMNS matrix.
The fit also gives $ \delta^\ell \approx -120^\circ $, in agreement with a previous numerical study of CSD($n$) models \cite{bjorkeroth14}, and encouragingly close to current experimental hints for a normal neutrino hierarchy.
However, phase freedom in the mass matrices allows also a fit where $ \delta^\ell \approx +120^\circ $, with other parameters essentially the same.
CSD(4) alone cannot predict the sign of $ \delta^\ell $.

\section{Flavour constraints on the flavourful axion scale}
\label{sec:axionflavour}

Contrary to the usual KSVZ or DFSZ axion model, a flavourful axion model allows general flavour-violating couplings of the axion 
which may constrain the axion scale more strongly. 
As noted in \cite{ema16},  a severe limit is obtained from the kaon decay $K^+ \to \pi^+ a$. 
The Yukawa structure in Eq.~\ref{Yuk}
leads to the mass matrix mis-aligned with the axion coupling matrix due to the flavour-dependent PQ charges.
As a result, our model predicts a specific flavour-violating coupling to down ($d$) and strange ($s$) quarks
\begin{equation} \label{asd}
	\mathcal{L}_{asd} = i \frac{a}{N_a f_a} \left[ \mathrm{Re}(m^d_{21}) \bar s \gamma_5 d + \mathrm{Im}(m^d_{21}) \bar s d\right] ,
\end{equation}
where $m^d_{21}=B |y^0_d| v_d/\sqrt{2}$. 
To understand the above equation, note that a flavon field $\varphi$ is effectively expressed by $\varphi= v_\varphi e^{i a_\varphi/v_\varphi}$ where  the phase field contains the axion component as $a_\varphi = x_\varphi v_\varphi a / v_{PQ} + \cdots$ (see Eq.~\ref{a-aphi}). 
Thus we get $\varphi = v_\varphi(1+ i x_\varphi a/v_{PQ} +\cdots)$ inducing the axion coupling matrix to down-type quarks
\begin{equation}
Y^d_{a} = \pmatr{
		2y_d^0  & 0 & 0 \\
		2B y_d^0  & y_s^0  & 0 \\
		2 B y_d^0  & 0 & 0
	}.
\end{equation}
Therefore, one finds the $a$-$s$-$d$ coupling of Eq.~\ref{asd} in the mass basis
after diagonalising $Y^d$ in Eq.~\ref{Yuk}.

The present experimental limit, B($K^+\to \pi^+ a) < 7.3\times 10^{-11}$ \cite{adler08},
puts the  bound of $\mbox{Im}(m^d_{21})/N_a f_a < 1.7 \times 10^{-13} $. One thus finds
\begin{equation}
	N_a f_a > 2.3 \times 10^{10} \mathrm{~GeV},
\end{equation}
taking the central values of our input parameters shown in Table 4. 
This tells us a rough bound on the flavon VEVs: $\braket{\phi^{u,d}_{1,2}} \gtrsim 10^{10}$ GeV.
The NA62 experiment is expected to reach the sensitivity of B($K^+\to \pi^+ a) < 1.0\times 10^{-12}$ \cite{na62},
probing $N_a f_a$ up to $2 \times 10^{11}$ GeV.

\section{Conclusion}
\label{sec:conclusion}

We have investigated the possibility that an accidental PQ symmetry could arise from discrete flavour symmetry,
which represents the first study of its kind. The ingredients of the model are a discrete flavour symmetry 
which encompasses both leptons and quarks, where the PQ symmetry is not imposed by hand but emerges accidentally,
and is spontaneously broken by flavons, resulting in a flavourful axion.

To be concrete, we have presented a solution to the strong $ CP $ problem in a supersymmetric unified model of flavour, 
which is a modification of the A to Z of flavour Pati-Salam model, based on Pati-Salam and $ A_4 $ symmetry,
together with an Abelian discrete flavour symmetry.
With some modifications to the original model, an accidental Peccei-Quinn symmetry is realised at the renormalisable level, and spontaneously broken by the VEVs of $ A_4 $ triplet flavons, which are also responsible for explaining the flavour structures of quarks and leptons via the CSD(4) vacuum alignments.
For the first time, we have shown how a PQ symmetry can arise purely from discrete flavour symmetries,
where we have ensured that the accidental $ U(1)_{PQ} $ is protected to sufficiently high order, with all higher-order operators suppressed by the Planck mass are forbidden up to dimension 10.

To achieve this, in addition to the original $ A_4 \times \mathbb{Z}_5 $ discrete flavour symmetry, we also
introduced a $ \mathbb{Z}_3 \times \mathbb{Z}_5^\prime $ symmetry, under which flavons as well as right-handed SM matter fields (contained in PS multiplets $ F^c_i $) are charged.
However, no new scalar fields were necessary to realise the $ U(1)_{PQ} $ symmetry: the same flavons already introduced to explain flavour structures also give rise to a QCD axion.
The $ \mathbb{Z}_3 $ symmetry is sufficient to ensure the PQ symmetry at the renormalisable level.
It also necessarily forbids several terms in the Yukawa superpotential allowed in the original model, modifying the predictions for flavour. 
The $ \mathbb{Z}_5 $ is primarily responsible for protecting the PQ solution against higher-order terms suppressed by powers of $ M_P $, up to $ D = 10 $.

The originally proposed superpotential which drives the flavons to have non-zero VEVs, commonly used to drive VEVs
in models of this kind,
turned out to be generally incompatible with a PQ symmetry.
In order to overcome this we have suggested an alternative mechanism, which respects the PQ solution, wherein flavons $ \phi $ couple to ``conjugate'' flavons $ \overbar{\phi} $ which have opposite charges under all symmetries.
The $ \mathbb{Z}_5^\prime $ symmetry is then essential also in forbidding dangerous renormalisable couplings of conjugate flavons to matter, which spoils both the PQ symmetry and the predictive flavour structures.

The accidental QCD axion arising from the flavourful PQ symmetry in our model shares many phenomenological properties 
with the conventional DFSZ axion. A crucial difference comes from its flavour-violating couplings determined by the predicted
Yukawa structure and flavour-dependent PQ charges. The specific prediction for the $a$-$s$-$d$ coupling allows us to probe 
the axion scale $f_a$ up to $3 \times 10^{10}$ GeV in the NA62 experiment.
We look forward to a new era in flavourful axion model building and phenomenology,
where the discrete symmetries responsible for flavour may also accidentally yield a global PQ symmetry and 
resolve the strong $CP$ problem.

\subsection*{Acknowledgements}
SFK acknowledges the STFC Consolidated Grant ST/L000296/1 and the European Union's Horizon 2020 Research and Innovation programme under Marie Sk\l{}odowska-Curie grant agreements Elusives ITN No.\ 674896 and InvisiblesPlus RISE No.\ 690575. 
EJC is supported also by InvisiblesPlus RISE No.\ 690575.
FB~is supported in part by the INFN ``Iniziativa Specifica'' TAsP-LNF. 
FB also thanks colleagues at KIAS for their hospitality during the development of this work.

\end{document}